\documentclass{article}
\usepackage[T1]{fontenc}
\usepackage{amsmath}
\usepackage{amssymb}
\usepackage{textcomp}
\usepackage{listings}
\usepackage{verbatim}
\usepackage{xcolor}
\usepackage{authblk}
\usepackage[margin=1.25in]{geometry}
\usepackage{xspace}

\newcommand{\kconfig}{Kconfig\xspace}

\newcommand{\Id}{\textsf{Id}}

\newcommand{\Type}{\textsf{Type}}
\newcommand{\Default}{\textsf{Default}}
\newcommand{\Range}{\textsf{Range}}
\newcommand{\KConfig}{\textsf{Kconfig}}
\newcommand{\Configs}{\textsf{Configs}}

\newcommand{\Choices}{\textsf{Choices}}
\newcommand{\KExpr}{\textit{KExpr}}
\newcommand{\Powerset}[1]{\ensuremath{\mathcal{P}(#1)}}
\newcommand{\Confs}{\textsf{Confs}}
\newcommand{\Lower}{\textsf{Lower}}
\newcommand{\Upper}{\textsf{Upper}}
\newcommand{\Bool}{\textsf{Bool}}
\newcommand{\Tri}{\textsf{Tri}}
\newcommand{\String}{\textsf{String}}
\newcommand{\Int}{\textsf{Int}}
\newcommand{\Hex}{\textsf{Hex}}

\newcommand{\Const}{\textsf{Const}}

\newcommand{\bool}{\ensuremath{\mathrm{bool}}}
\newcommand{\eval}{\textit{eval}}
\newcommand{\ndef}{\textit{defs}}

\newcommand{\mconfig}{\ensuremath{m_\mathsf{config}}}
\newcommand{\mchoice}{\ensuremath{m_\mathsf{choice}}}

\newcommand{\kboolean}{\textrm{boolean}}
\newcommand{\ktristate}{\textrm{tristate}}
\newcommand{\kstring}{\textrm{string}}
\newcommand{\khex}{\textrm{hex}}
\newcommand{\kint}{\textrm{int}}

\newcommand{\denotation}[1]{\ensuremath{[\![#1]\!]}}
\newcommand{\semkconfig}[1]{\ensuremath{\denotation{#1}_{\textsf{kconfig}}}}
\newcommand{\sembounds}[1]{\denotation{#1}_{\textsf{bounds}}}
\newcommand{\semdefault}[1]{\denotation{#1}_{\textsf{default}}}
\newcommand{\semchoice}[1]{\denotation{#1}_{\textsf{choice}}}
\newcommand{\semtype}[1]{\denotation{#1}_{\textsf{type}}}
\newcommand{\semrange}[1]{\denotation{#1}_{\textsf{range}}}
\newcommand{\semmod}[1]{\denotation{#1}_{\textsf{module}}}
\newcommand{\semundec}[1]{\denotation{#1}_{\textsf{undeclared}}}

\newcommand{\tri}[1]{\ensuremath{#1_\mathsf{t}}}

\DeclareMathOperator{\mnot}{not}
\DeclareMathOperator{\mor}{\mathbin{or}}

\DeclareMathOperator{\mand}{\mathbin{and}}
\DeclareMathOperator{\mneq}{\not =}

\title{Formal Semantics of the Kconfig Language\\[.2\baselineskip]\normalsize\textbf{Technical Note}}

\author[1]{Steven She}
\affil[1]{\textsf{shshe@gsd.uwaterloo.ca}\\University of Waterloo, Canada}
\author[2]{Thorsten Berger}
\affil[2]{\textsf{berger@informatik.uni-leipzig.de}\\University of Leipzig, Germany}
\date{January 2010}
\begin{document}

\maketitle

\begin{abstract}
  The \kconfig language defines a set of symbols that are assigned a value in a configuration. We describe the semantics of the \kconfig language according to the behaviour exhibited in the xconfig configurator.  We assume an abstract syntax representation for concepts in the \kconfig language and delegate the details of the translation from concrete to abstract syntaxes to a later document.
\end{abstract}

\footnotetext{The semantics defined in this document directly reflect the behavior of the Linux \textit{make xconfig} tool, which could \textthreequartersemdash\ in some specific cases \textthreequartersemdash\ act differently from what the \KConfig\ language developers originally had in mind. At least in case of the \textit{reverse dependency} the documentation explicitly states the following gap: "\textit{Select} should be used with care. \textit{Select} will force a symbol to a value without visiting the dependencies. By abusing \textit{select} you are able to select a symbol \textsc{foo} even if \textsc{foo} depends on \textsc{bar} that is not set."}

\section{Abstract syntax}

\paragraph{Identifiers and expressions.} We start be defining the preliminary concepts available in the \kconfig language. Let \Id\ be a finite set of names identifying a symbol---more precisely, $\Id \in \Powerset\String$. Let $\Const = \Tri \cup \String \cup \Hex \cup \Int$ be the set of values assignable to each feature and available as constants in expressions, where $\Tri = \{ \tri0, \tri1, \tri2 \}$. \Tri\ is ordered such that $\tri0 < \tri1 < \tri2$. The \Tri, \String, \Hex, and \Int\ domains are disjoint (i.e.\ mutually exclusive). We can now define an expression in the \kconfig language. $\KExpr(\Id)$ is a set of expressions over \Id\ generated by the following grammar, where $e \in \KExpr(\Id)$, $iv \in \Id \cup \Const$, $\otimes\in\{\mor, \mand\}$, $\ominus \in \{ =, \mneq \}$:
\begin{equation}
  e ::= e \otimes e 
  \mid \mnot e 
  \mid iv \ominus iv 
  \mid iv
\end{equation}

Evaluating a $\KExpr$ returns a tristate value (i.e.\ $v \in \Tri$). We will define the semantics of an $eval$ function in Section~\ref{sec:eval}. 

\paragraph{\kconfig model.} \KConfig\ denotes the set of all possible models in the Kconfig language.  Thus a single \kconfig model $m\in\KConfig$ is a tuple consisting of a set of \textit{configs} and a set of \textit{choices}. \KConfig\ is defined as:

\begin{equation}
  \KConfig = \Powerset{\Configs} \times \Powerset{\Choices}
\end{equation}

\noindent Given a \kconfig model $m \in \KConfig$, we define the shorthand \mconfig\ to refer to its set of configs and \mchoice\ to refer to its set of choices.

Configs are the primary components of a \kconfig model. A config defines a unique identifier with type, a prompt condition --- a condition that determines when a config becomes user-changeable, a list of defaults, a expression denoting its reverse dependency --- the conditions that would forcefully enable this feature through a \textit{select} statement, and a set of ranges --- restrictions on the value for configs or hex type. We define \Configs as follows:
\begin{equation}
	\Configs = \Id \times \Type \times \KExpr(\Id) \times \Default* \times \KExpr(\Id) \times \Powerset\Range
\end{equation}

\begin{samepage}
\noindent where,
\begin{itemize}
  \item $\Type=\{\text{boolean},\text{tristate},\text{int},\text{hex},\text{string}\}$ denotes a type and consequently the possible values for the config.
  \item $\Default = \KExpr(\Id)\times\KExpr(\Id)$ denotes defaults. The first \KExpr denotes a default expression (i.e.\ that is evaluated and assigned to the symbol) and the second \KExpr denotes the condition required for the default to become effective.
  \item $\Range = (\Int \cup \Hex \cup \Id) \times (\Int \cup \Hex \cup \Id) \times \KExpr(\Id)$ is a triple consisting of a lower bound, an upper bound and a condition. Note the absence of \Tri\ in the lower and upper bounds of the range; this is due to ranges being only effective on int and hex-typed configs as we will describe in the following paragraph.
\end{itemize}
\end{samepage}

We further define a function $\Id(m)$ to denote identifiers of configs in the model $m$:
\begin{equation}
\Id(m) = \{n \mid (n,\_,\_,\_,\_,\_)\in \mconfig \}
\end{equation}

The second component of \KConfig\ refers to a set of choice nodes. A choice is an abstract construct that defines no symbol in the configuration, however, it imposes additional constraints on its nested elements. We define choices as a quadruple consisting of a type where boolean or tristate are the only valid types, a flag indicating whether the choice is mandatory, a prompt condition followed by a set of identifiers indicating its members. The set $\Choices$ is defined as:

\begin{equation}
  \Choices = \{ \kboolean, \ktristate \} \times \Bool \times \KExpr(\Id) \times \Powerset{\Id(m)}
\end{equation}

\paragraph{Well-formedness rules.} Given an element $(\_, t, \_, \_, rev, rngs) \in \Configs$, this config is well-formed if the following conditions are satisfied:

\begin{itemize}
  \item The reverse dependency of configs with type int, hex or string must be $\tri0$. In other words, no config may select a config that is not of type boolean or tristate.
    \begin{equation}
      (rev \neq \tri0) \implies t = \kboolean \lor t = \ktristate
    \end{equation}

  \item Ranges can only be defined on configs with a numerical type, namely int or hex types. Thus, the following constraint must hold for a config to be well-formed: 
    \begin{equation}
    (|rngs| > 0) \implies t = \kint \lor t = \khex
    \end{equation}

\end{itemize}

\paragraph{Brief note on concrete syntax translation.}
\textit{Menuconfigs} and \textit{menus} are first-class concepts in the concrete syntax of the \kconfig language. However, both of these concepts are not present in the abstract syntax. First, menuconfigs are semantically identical to configs and only differ in terms of its appearance in the configurator; thus, we model menuconfigs as configs in the abstract syntax. Menus do not define a symbol; thus menus are not present in a configuration. However, menus can impose constraints on its nested elements. We handle these constraints via a syntactic rewrite on the prompt, default and range conditions of all nested symbols. Details for this syntactic rewrite will be provided in later document.

\section{Semantics}

\subsection{Semantic domain}
A configuration of a \KConfig\ model is an assignment of values $v\in\Const$ to config elements. Thus, the set of all possible configurations is defined as:
\begin{equation}
\Confs=\Id \rightarrow \lfloor\Const\rfloor
\end{equation}

If $c\in\Confs$ and $x\in\Id$, we write $c(x)$ in order to refer to the value of identifier $x$ under the configuration $c$. Now, we define the semantics of a \KConfig\ model in terms of sets of configurations. Thus, \Powerset\Confs\ is our semantic domain. We define $\semkconfig\cdot$ as the function that evaluates a \kconfig model and returns a set of valid configurations:
\begin{equation}
\semkconfig\cdot \colon \KConfig\rightarrow\Powerset\Confs
\end{equation}

\subsection{Global functions}
\label{sec:eval}
We start with the definition of some functions used throughout the semantics. First, we define an interpretation of tristate values in boolean logic with $\bool\colon \Tri \rightarrow \Bool$ where $\Bool = \{ T, F \}$:
\begin{equation}
\bool( v ) = 
	\begin{cases}
		F & \text{iff~} v=0_t\\
		T & \text{iff~} v=1_t \vee v=2_t
	\end{cases}
\end{equation}

Moreover, we define a function $access:(\Id\cup\Const)\times\Confs\rightarrow\Const$ that retrieves the value of either a constant or a symbol. When an identifier has the value of $\bot$ (to be defined in Equation~\ref{eqn:kconf}), then the $access$ function returns the identifier itself in the form of a string:
\begin{equation}
\mathrm{access}(iv,c)=\begin{cases}
iv & \text{iff~} iv\in\Const \vee ( iv\in\Id \land c(iv) = \bot ) \\
c(iv) & \text{otherwise}\\
\end{cases}
\end{equation}

Next, we define the function $toStr\colon \Const \rightarrow \String$ that models the translation of a constant to a string representation. Let $i \in \Int$, $h \in \Hex$  and $s \in \String$, in the following definition of $toStr$:
\begin{equation}
  \begin{aligned}
    toStr(\tri0) &= \text{``n''} & toStr(\tri1) &= \text{``m''} & toStr(\tri2) &= \text{``y''} \\
    toStr(i) &= \text{``''} + i & toStr(h) &= \text{``0x''} + h & toStr(s) &= s
  \end{aligned}
\end{equation}

\noindent where the $+$ operator is string concatenation.

Finally, the function $eval\colon KExpr(\Id) \rightarrow \Tri$ describes the evaluation of a $KExpr$ in the \kconfig language. We define $eval$ recursively with $e_1, e_2 \in KExpr(\Id)$ and $iv, iv_x, iv_y \in \Id \cup \Const$:
\begin{equation}
\begin{split}
  \eval (iv_x = iv_y, c) &=  
  \begin{cases}
    \tri2\ & \text{iff~} toStr(\textrm{access}(iv_x,c)) = toStr(\textrm{access}(iv_y,c)) \\
    \tri0 & \text{otherwise}
  \end{cases} \\
  \eval (iv_x \mneq iv_y, c) &= \tri2 - \eval(iv_x = iv_y, c) \\
  \eval(\mnot e_1, c) & =  \tri2 - eval(e_1 ,c) \\
  \eval(e_1 \mand e_2, c) & = \textrm{min}(eval(e_1,c), eval(e_2,c)) \\
  \eval(e_1 \mor e_2, c) &=  \textrm{max}(eval(e_1,c), eval(e_2,c)) \\
  \eval(iv,c) & = 
  \begin{cases}
    v_{iv} & \text{iff~} v_{iv}=\textrm{access}(iv,c) \land v_{iv} \in \Tri\\
    \tri0 & \text{otherwise}
  \end{cases}
\end{split}
\end{equation}

\subsection{Valuation functions} 
\paragraph{\kconfig model.} We begin by defining the $\semkconfig\cdot$. Given a \kconfig model $m \in \KConfig$, the semantics of a model is the intersection of all denotations across the model, configs and choices.
In other words, the set of valid configurations for a \kconfig model is those configurations that satisfy all denotations. $\semkconfig\cdot \colon \KConfig \rightarrow \Confs$ is defined:

\begin{equation}\label{eqn:kconf}\begin{split}
\semkconfig m =&
\left(\bigcap_{n\in \mconfig} \semtype n \cap \sembounds n \cap \semdefault n \cap \semrange n \right) \cap \left( \bigcap_{n\in \mchoice}\semchoice n \right) \\
& \qquad\cap \semmod m \\
& \qquad\cap \semundec m
\end{split}
\end{equation}

\paragraph{Type.} The first denotation pertains to the constraints imposed by a config's type.   The type of a config restricts its valid values to those in its respective domain.  $\semtype\cdot \colon \Configs \rightarrow \Confs$ is defined:
\begin{equation}
  \semtype{(n,t,\_,\_,\_,\_)} =
  \begin{cases}
    \left\{ c \in \Confs \mid c(n) \in \Tri \setminus \{\tri1\} \right\}  & \text{iff } t = \kboolean \\
    \left\{ c \in \Confs \mid c(n) \in \Tri \right\} & \text{iff } t = \ktristate \\
    \left\{ c \in \Confs \mid c(n) \in \String \right\} & \text{iff } t = \kstring \\
    \left\{ c \in \Confs \mid c(n) \in \Hex \cup \{ \text{``''} \} \right\} & \text{iff } t = \khex \\
    \left\{ c \in \Confs \mid c(n) \in \Int \cup \{ \text{``''} \} \right\} & \text{iff } t = \kint
  \end{cases}
\end{equation}

\paragraph{Upper and lower bounds.} Next, the bounds denotation models the lower and upper bounds of a config. The lower bound is determined by the evaluation of a config's reverse dependency. Recall that the reverse dependency models the behaviour of the \textit{select} statement in the concrete syntax. The upper bound is defined by a config's prompt condition. This denotation has no effect on configs of type int, hex, or string since the the reverse dependency that determines a lower bound is $\tri0$ by our well-formedness rules, and the $eval$ function returns $\tri0$ when evaluating a value not in $\Tri$. $\sembounds\cdot \colon \Configs \rightarrow \Confs$ is defined:
\begin{equation}
\begin{split}
\sembounds{(n,\_,&\,pro,\_,rev,\_)} = \\
& \left\{ c\in\Confs\mid eval(c(n),c) \geq \Lower(c) \wedge (\Upper(c) < \Lower(c) \vee eval(c(n),c) \leq\Upper )\right\}
\end{split}
\end{equation}

\noindent where \Lower(c) = \eval(rev,c) and \Upper(c) = \eval(pro,c).

\paragraph{Defaults.} \kconfig has support for setting a default expression for a config. The default expression interacts with the prompt condition that determines when the config is user-changeable. When the prompt condition is satisfied, then the user is free to set a value. However, when the prompt condition is not satisfied, the default determine the config's value. $\semdefault\cdot \colon \Configs \rightarrow \Confs$ is defined:
\begin{equation}
\begin{split}
\semdefault{(n,\_,&\,\_,defs,rev,\_)} = \\
&\left\{ c\in\Confs\mid \bool(\eval(pro,c)) \vee c(n)= \mathrm{max}( eval(\mathrm{default}(\ndef,c)), eval(rev,c) ) \right\}
\end{split}
\end{equation}

\noindent where $default\colon \Powerset\Default \times \Type \times \Confs \rightarrow \Const$ is a function that models the retrieval of a default. Recall that $defs$ is a list of defaults (and thus ordered). The effect of a default's value depends on the type of its defining config. If the config is $boolean$ or $tristate$, then the default value is evaluated to a value in \Tri. Otherwise, the default value must be either an element of \Const\ or \Id. Let $Nil$ be the empty list and $::$ be the list $cons$ operator. Let $t_\Tri \in \{ \kboolean, \ktristate \}$ and $t_\mathsf{Entry} \in \{ \kint, \khex, \kstring \}$. The $default$ function is defined recursively, so we begin by defining its base cases:
\begin{equation}\label{eq:defbase}
  \begin{aligned}
    \mathrm{default}(Nil,t_\Tri,c) &= \tri0 \\
    \mathrm{default}(Nil,t_\mathsf{Entry},c) &= \text{``\,''}
  \end{aligned}
\end{equation}

\noindent Equation~\ref{eq:defbase} states that given an empty list of defaults, we return $\tri0$ if the type is either boolean or tristate, or the empty string for types int, hex or string. Next, we define the recursive rule. In the following equation, we decompose the list into its head and tail components. First, we describe the function for boolean and tristate type:
\begin{equation}
  \mathrm{default}(\, (e, cond) :: rest,t_\Tri,c) = 
      \begin{cases}
        eval(e, c) & \text{if~} \mathrm{bool}(eval(cond,c)) \\
        \mathrm{default}(rest,t_\Tri,c) & \text{otherwise}
      \end{cases}
\end{equation}

\noindent Now for the remaining types:
\begin{equation}
  \mathrm{default}(\, (e, cond) :: rest,t_\mathsf{Entry},c) = 
      \begin{cases}
        \mathrm{access}(e, c) & \text{if~} \mathrm{bool}(eval(cond,c)) \\
        \mathrm{default}(rest,t_\mathsf{Entry},c) & \text{otherwise}
      \end{cases}
\end{equation}

\paragraph{Ranges.} Ranges impose a lower and upper bound on the value of int or hex configs. $\semrange\cdot \colon \Configs \rightarrow \Confs$ is defined as:
\begin{multline}
  \semrange{n,\_,\_,\_,\_,rngs)} = \{ c \in \Confs \mid\ \forall (l,u,cond) \in rngs.\\
  \mathrm{bool}(eval(cond,c)) \rightarrow c(n) \geq \mathrm{access}(l,c) \land c(n) \leq \mathrm{access}(u,c) \}
\end{multline}

\paragraph{Choices.} A choice restricts the number of members that can be selected (i.e.\ have a value greater than $\tri0$). The choice denotation, $\semchoice\cdot \colon \Choices \rightarrow \Confs$ is defined: 
  \begin{multline}
  \semchoice{(boolOrTri, isMand, prompt, mems)} = \\
  \{ c \in \Confs \mid\ bool(eval(prompt,c)) \rightarrow \textsf{Xor} \land \textsf{BChoice} \land \textsf{Mandatory} \}
  \end{multline}

  \noindent where \textsf{Xor} defines the condition that one and only one member may be set to $\tri2$:
\begin{equation}
  \textsf{Xor} = \exists m_1 \in mems.\ (m_1 = \tri2) \rightarrow (\forall m_2 \in mems \setminus \{ m_1 \}.\ m_2 = \tri0)
\end{equation}

\noindent If the choice is a $\kboolean$ choice, then the only valid value for its members is $2_t$. In combination with \textsf{Xor}, this defines that a $\kboolean$ choice may have at most one member with a value not equal to $\tri0$ and that member must be set to $\tri2$:
\begin{equation}
  \textsf{BChoice} =  (boolOrTri = \kboolean) \rightarrow \exists m \in mems.\ c(m) = \tri2
\end{equation}

\noindent Finally, if the choice is \textit{mandatory}, then at least one member must be selected:
\begin{equation}
  \textsf{Mandatory} = isMand \rightarrow \exists m \in mems.\ c(m) > \tri0
\end{equation}

\paragraph{Modules.} A special \textsc{modules} config is used to specify support for modules in the kernel. Disabling \textsc{modules} disallows the $1_t$ state for configs and effectively turns all tristate configs into boolean configs. A special symbol $m$ is used in expressions to identify a dependency on the \textsc{modules} feature in the concrete syntax. Configs with a dependency on $m$ cannot be selected (i.e.\ must be set to $0_t$) if \textsc{modules} is not selected. We assume that the special $m$ identifier has been expanded to \textsc{modules} in the abstract syntax. 
\begin{equation}
  \semmod{m} = \{ c \in \Confs \mid c(\mathrm{\textsc{modules}}) = \mathrm{n} \rightarrow \forall i \in \Id.\ c(i) \neq 1_t \}
\end{equation}

\paragraph{Undeclared symbols.} We also define the behaviour of \textit{undeclared symbols}. The \kconfig language supports references to symbols that are not declared in constraints. These undeclared symbols are assigned the special symbol $\bot$ in our semantics. The use of this symbol will become apparent in the definition of the $eval$ function in Section~\ref{sec:eval}. The $\semundec\cdot \colon \KConfig \rightarrow \Powerset\Confs$ denotation is defined as:
\begin{equation}
  \semundec{m} = \{c\in\Confs\mid \forall x\in\Id\setminus\Id(m).\ c(x)=\bot\}
\end{equation}

\section{1-Var Propositional Semantics}

The goal of the propositional semantics is to achieve a weakening of the constraints of the full semantics.

\begin{table}
\centering
\begin{tabular}{rll}
  & type & interpretation in tristate logic \\
  \hline
  $X$ & tristate & $X = y$ or $X = m$ \\
  $\lnot X$ & tristate & $X = n$ \\
  $X$ & boolean  & $X = y$ \\
  $\lnot X$ & boolean & $X = n$ \\
  $X$ & string & $X = \text{``\dots''}$ (some non-empty string) \\
  $\lnot X$ & string & $X = \text{``\,''}$ \\
  $X$ & int & $X = i$ (some integer, including 0) \\
  $\lnot X$ & int & $X = \text{``\,''}$ \\
  $X$ & hex & $X = i$ (some hex) \\
  $\lnot X$ & hex & $X = \text{``\,''}$ \\
\end{tabular}
\caption{\label{tab:interp}Interpretation of propositional variables}
\end{table}

\paragraph{Rewrite rules for expressions.}

The $rewrite$ is a partial function that implements rewrite rules on expressions. The function $rewrite\colon \KExpr(\Id) \rightarrow \KExpr(\Id)$ is defined as:                                  
\begin{equation}
    rewrite(e) =
    \begin{cases}
      0 & \text{if~} (e \text{ is an variable } \land typeOf(e) \in \{ int, hex, string \}) \lor e = 0_t\\
      1 & \text{if~} e = 1_t \lor e = 2_t \\
      X \leftrightarrow Y & \text{if~} e \text{ is } X = Y \text{ where X and Y are variables} \\
      \text{\textit{Use Table~\ref{tab:interp}}} & \text{if~} e \text{ is } X = lit \lor X \neq lit\\
    \end{cases}
\end{equation}

We further define the function $relax\colon \KExpr(\Id) \rightarrow \KExpr(\Id)$ which converts an expression to CNF and removes clauses equivalent to equality checks. This function is used to relax the constraints on the antecedent (LHS) of an implication.

\paragraph{Semantics.}
In the propositional semantics, we model the set of propositional configurations as $\Confs_p$:
\begin{equation}
  \Confs_p = \Id \rightarrow \Bool
\end{equation}

The $default$ function is defined as $\Id \times \Default* \times \Confs_p \rightarrow \Bool$. An extra parameter providing the declaring identifier is needed for the propositional semantics.

\begin{equation}
  default(n,defs,c) =
  \begin{cases}
    \lnot n & \text{if no default conditions are satisfied} \\
    rewrite(n = eval(iv_i,c)) & \text{otherwise if~} t \in \{ boolean, tristate \}, \\
    &\text{where } iv_i \text{ is the 1st matching default value}  \\
    n & \text{otherwise if~} t \in \{ int, hex, string \} \\
  \end{cases}
\end{equation}

The default denotation is defined as:
\begin{equation}
  \semdefault{(n,t,vis,pro,defs,rev,rngs)} = \{ c \in \Confs_p \mid eval(pro,c) \lor default(n,defs,c) \}
\end{equation}

The constraint denotation which models constraints imposed by the reverse dependency and visibility conditions is defined as:
\begin{equation}
  \sembounds{(n,t,vis,pro,defs,rev,rngs)} = \{ c \in \Confs_p \mid (eval(relax(rev),c) \rightarrow c(n)) \land (c(n) \rightarrow eval(vis,c)) \}
\end{equation}

We ignore ranges since we abstract away the value of each config. We also assume that the \textsc{module} config is enabled, thus allowing for the $1_t$ state in tristate configs.

\begin{multline}
  \semchoice{(boolOrTri, isMand, vis, mems)} = \\
  \left\{ c \in \Confs \mid eval(vis,c) \rightarrow choose(1, ids(mems)) \land \left(isMand \rightarrow \bigvee_{m \in mems} m\right)\right\}
\end{multline}








\end{document}